# Classical treatment of evanescent electromagnetic fields interacting with an amplifying or attenuating medium upon total internal reflection

**Takeyuki Kobayashi**

Department of Opto-Electronic System Engineering, Chitose Institute of Science and Technology, 758-65 Bibi, Chitose, Hokkaido 066-8655, Japan

The present study deals with total internal reflection of a plane electromagnetic wave at an infinite plane boundary between a transparent medium and an amplifying or attenuating lower-index medium. Solutions of Maxwell's equations are sought with the effect of gain or loss taken into account by means of the effective indices of refraction and extinction associated with the real angles representing the directions of propagation and amplitude change. Analytical expressions are found for the reflection coefficients and the refracted electric and magnetic fields, which describe the amplification or attenuation of the evanescent field resulting from the interaction with the totally reflecting medium.

# I. Introduction

There exists a class of electromagnetic waves termed inhomogeneous, for which the direction of propagation is, in general, do not coincide with that of amplitude change, unlike homogeneous waves whose directions of propagation and of amplitude change coincide with each other. Among various physical situations which may give rise to such inhomogeneous waves, total internal reflection has been a topic of particular interest urged by the need of understanding the interaction of the evanescent inhomogeneous wave with the totally reflecting medium exhibiting either gain or loss.

Over the years there have been experimental and theoretical studies to gain insight into the interaction of the evanescent wave with an amplifying or attenuating medium. The first recorded attempt dates back to the clever demonstration by Selenyi [1], where luminescent molecules placed on one plane of a prism were excited by the evanescent field when the light propagating inside the prism was totally reflected at its inner surface.

The pioneering work by Carniglia *et al*. [2] has shown that, upon total reflection, the incoming plane wave is absorbed as a result of the interaction of the evanescent wave with the absorbing molecules in the low-index medium and also that an evanescent wave can be excited by spontaneous radiation from the luminescent molecules in the vicinity of the boundary plane, and then can be coupled out in the form of homogeneous plane wave. In [3, 4] are presented the results of theoretical investigation on the interaction of the evanescent wave with the low-index attenuating medium.

The observation of optical gain by stimulated emission [5] added an intriguing prospect because, instead of the attenuation resulting from losses due to absorption or scattering in the totally-reflecting medium, there emerged the possibility of evanescent-field amplification resulting in a greater-than-unity reflection coefficient. Indeed, the proposal was made of amplifying the guided modes interacting with a cladding with gain [6].

The idea has since been pursued both experimentally and theoretically. A few different terms have been used for this phenomenon such as evanescent gain, amplified total internal reflection, enhanced total reflection, and others. Some work tried to quantify the magnitude of a single reflection [7, 8] whereas the other measured the result of accumulation of multiple reflection employing some structure supporting guided modes [9-12].

As for theoretical work performed to date on evanescent gain [13-17], both the methods and the predicted values of the reflection coefficient are rather diverse, hardly leading us to a unanimous conclusion. The work by Lee *et al.* asserts that a quantum-mechanical treatment is required to settle the problem [18]. Contrary to the theoretical studies cited above, Siegman argued that there does not exist a solution to Maxwell's equations, which describes evanescent gain [19]. In [20], the coefficient for a single reflection was calculated by working backwards from the amplitude change of a waveguide mode resulting from the accumulated multiple reflection. The expressions obtained there are valid for both gain and loss.

The present study is an attempt to answer the long-standing problem of evanescent gain and loss. In the following two sections, II and III, the concepts of the effective indices of refraction and extinction are introduced, which allow us to define the phase and amplitude propagation vectors, the useful tools to describe the behaviour of the refracted waves in a medium having gain or loss. The *s*-polarisation case is treated throughout the main body of the paper. Along with the physical interpretations of the derived expressions, the formulation of Snell's law is discussed. Section IV deals with the phase change upon total reflection with the effect of the index of extinction taken into account. Section V is concerned with the amplitude reflection coefficient, where explicit expressions for the electric and magnetic fields of the evanescent wave are given. Then comes the concluding section, to be followed by an appendix, which provides the formulas for the *p*-polarisation case.

## II. Effective indices of refraction and extinction

We begin by establishing our coordinate system and the notation for the relevant physical parameters. A plane boundary of two adjoining homogeneous and isotropic media, each filling half space, is assumed. One of the two media is lossless and transparent and the other with a lower refractive index has either gain or loss. Both of them are nonmagnetic. The boundary is represented by the plane $z = 0$. We consider a monochromatic infinite plane wave with a harmonic time dependence exp($j\omega t$), which is incident on the boundary from the transparent side at an angle $\theta$ from the normal to the plane, as shown in Fig. 1. Here we shall confine ourselves to the *s*-polarisation case, where the electric field is perpendicular to the plane of incidence $y = 0$ and leave the *p*-polarisation case to be treated in the appendix.

As a starting point of our analysis, we lay out the expressions for the incident, reflected, and refracted waves when the reflecting medium is transparent ($\kappa = 0$). Let the incident electric field be represented by the real parts of the following expression

$$E_y (x, z, t) = E_s \exp[-j(\omega/c)n_h(x \sin \theta + z \cos \theta)]. \tag{2.1}$$

Here *c* is the speed of light in vacuum. The time dependence exp($j\omega t$) is dropped as understood. Using the Maxwell equations satisfied by the *s*-polarised waves, we have the associated magnetic field

$$H_x (x, z, t) = (k_0 n_h/\mu_0 \omega) \cos \theta \, E_s \exp[-j(\omega/c)n_h(x \sin \theta + z \cos \theta)] \tag{2.2}$$

and

$$H_z (x, z, t) = (-k_0 n_h/\mu_0 \omega) \sin \theta \, E_s \exp[-j(\omega/c)n_h(x \sin \theta + z \cos \theta)]. \tag{2.3}$$

The incident wave gives rise to a reflected plane wave going off to the left of the boundary plane at an angle of reflection $\theta$. The electric and magnetic components are given by

$$E_y' (x, z, t) = E_s' \exp[-j(\omega/c)n_h(x \sin \theta - z \cos \theta)], \tag{2.4}$$

$$H_x' (x, z, t) = (-k_0 n_h/\mu_0 \omega) \cos \theta \, E_s' \exp[-j(\omega/c)n_h(x \sin \theta - z \cos \theta)], \tag{2.5}$$

and

$$H_z' (x, z, t) = (-k_0 n_h/\mu_0 \omega) \sin \theta \, E_s' \exp[-j(\omega/c)n_h(x \sin \theta - z \cos \theta)], \tag{2.6}$$

On the other side of the boundary, $z > 0$, there emerges a refracted wave of the form:

$$E_y'' (x, z, t) = E_s'' \exp[-j(\omega/c)n_l(x \sin \psi + z \cos \psi)], \qquad (2.7)$$

$$H_x'' (x, z, t) = (-k_0 n_l/\mu_0 \omega) \cos \psi \, E_s'' \exp[-j(\omega/c)n_l(x \sin \psi + z \cos \psi)] \qquad (2.8)$$

and

$$H_z'' (x, z, t) = (-k_0 n_l/\mu_0 \omega) \sin \psi \, E_s'' \exp[-j(\omega/c)n_l(x \sin \psi + z \cos \psi)]. \qquad (2.9)$$

The angle of refraction is denoted by $\psi$ for $\kappa = 0$ instead of $\Psi_N$ as indicated in Fig. 1.

If $n_h \geq n_l$, there exists a certain angle of incidence $\theta_c$ at which the angle of refraction reaches $\pi/2$. One way of dealing with such an inhomogeneous refracted field that occurs for $\theta > \theta_c$ is to allow the angle of refraction $\psi$ to be complex $\psi - j\psi'$ [21]. Substituting $\psi - j\psi'$ with $\psi = \pi/2$ in the phase factor of (2.7), we have

$$\exp[-j(\omega/c)n_l \, x \sin \psi] \exp[-(\omega/c)n_l \, z \sinh \psi'] = \exp[-j(\mathbf{k}_{l0} - j\boldsymbol{\alpha}_{\perp 0})\cdot \mathbf{r}],$$

where the phase propagation vector defined as $\mathbf{k}_{l0} = ((\omega/c)n_l \sin \psi, 0, 0)$ is parallel to the $x$-axis along the boundary plane and the amplitude vector $\boldsymbol{\alpha}_{\perp 0} = (0, 0, (\omega/c)n_l \sinh \psi')$ is perpendicular to the $x$-axis, hence to $\mathbf{k}_{l0}$ as well. The subscript 0 indicates the transparent case ($\kappa = 0$) throughout this paper. The $z$-dependence of the phase factor shows an amplitude profile that decays exponentially perpendicular to the boundary plane. The penetration depth, at which the amplitude of the evanescent field falls off to $1/e$ of that at $z = 0$, is defined as $d_{s0} = [\boldsymbol{\alpha}_{\perp 0}]_z^{-1} \, [= [(\omega/c)n_l \sinh \psi']^{-1} = (c/\omega)(n_h^2 \sin^2\theta - n_l^2)^{-1/2}]$. For later convenience, we show that by simple calculation the complex propagation vector is seen to satisfy the following relation:

$$(\mathbf{k}_{l0} - j\boldsymbol{\alpha}_{\perp 0})\cdot(\mathbf{k}_{l0} - j\boldsymbol{\alpha}_{\perp 0}) = (\omega/c)^2 n_l^2. \qquad (2.10)$$

To take into account the effect of amplification or attenuation in the classical theory, we replace the real index of refraction by a complex quantity with its imaginary part, which we shall call the index of extinction here, accounting for gain or loss. The low-index medium having a complex refractive index, the refracted wave becomes inhomogeneous at all angles of incidence except $\theta = 0$. In the optics literature, a formal substitution with a complex refractive index in Fresnel's formulas has always been assumed valid at all angles of incidence. Such an approach has indeed proven adequate under some circumstances, for example, when dealing with metallic reflection. As we shall see below, however, a different approach is required for the treatment of total internal reflection from complex-index media.

Starting from the phase of the refracted wave for the transparent case $n_l(x \sin \psi + z \cos \psi)$ in (2.7), we shall take up transforming the variables into those which are more amenable to analysing the propagation of inhomogeneous-waves. Since $\psi$ no longer represents the true angle of refraction, we shall define $\Psi_N$ and $\Psi_K$ together with the associated indices $N$ and $K$ by

$$(n_l - j\kappa) \sin (\psi - j\psi') = N \sin \Psi_N - jK \sin \Psi_K. \tag{2.11}$$

Equating the real and imaginary parts respectively, we have

$$N \sin \Psi_N = n_l \sin \psi \cosh \psi' - \kappa \cos \psi \sinh \psi' \tag{2.12}$$

and

$$K \sin \Psi_K = \kappa \sin \psi \cosh \psi' + n_l \cos \psi \sinh \psi'. \tag{2.13}$$

Thus, in an amplifying or attenuating medium, $\psi$ loses its physical meaning as the angle of refraction. Instead, $\Psi_N$ signifies the true angle of refraction.

Noting $(n_l - j\kappa) \cos(\psi - j\psi') = (+/-) (n_l - j\kappa) [1 - \sin^2(\psi - j\psi')]^{1/2}$, we have another pair of relations:

$$N \cos \Psi_N = (+/-)(n_l \cos\psi \cosh\psi' + \kappa \sin\psi \sinh\psi') \qquad (2.14)$$

and

$$K \cos \Psi_K = (+/-)(\kappa \cos\psi \cosh\psi' - n_l \sin\psi \sinh\psi'), \qquad (2.15)$$

where $N$ and $K$, which we shall call the effective indices of refraction and extinction, are obtained by eliminating $\Psi_N$ from (2.12) and (2.14) as

$$N = (n_l^2 \cosh^2\psi' + \kappa^2 \sinh^2\psi')^{1/2} \qquad (2.16)$$

and $\Psi_K$ from (2.13) and (2.15) as

$$K = (\kappa^2 \cosh^2\psi' + n_l^2 \sinh^2\psi')^{1/2}, \qquad (2.17)$$

respectively. Note that both $N$ and $K$ are not constant for a given set of $n_l$ and $\kappa$ but are dependent on the incident angle $\theta$ through $\psi'$.

The signs on the right-hand side of eqs. (2.14) and (2.15) should be chosen on physical grounds. To describe a wave travelling away from the boundary plane for $\Psi_N < \pi/2$, we take the plus sign in eq. (2.14). As for the sign in eq. (2.15), we shall have to exercise some care and discuss it at the end of Sec. III.

## III. Propagation vector for a complex-index medium

With the angles $\Psi_N$ and $\Psi_K$, along with the associated effective indices of refraction and extinction, $N$ and $K$, introduced in the preceding section, let us define a complex propagation vector $\mathbf{k}_l - j\boldsymbol{\alpha}_\perp$. Here $\mathbf{k}_l$ is the phase propagation vector given by

$$\mathbf{k}_l = (\omega/c)N (\sin \Psi_N, 0, \cos \Psi_N). \tag{3.1}$$

The imaginary component $\boldsymbol{\alpha}_\perp$, which we shall tentatively call the transverse amplitude vector, is perpendicular to the $\mathbf{k}_l$ vector, is given by

$$\boldsymbol{\alpha}_\perp = (\omega/c)K \cos (\Psi_N - \Psi_K) (-\cos \Psi_N, 0, \sin \Psi_N). \tag{3.2}$$

In addition to the above, we shall define the longitudinal amplitude vector $\boldsymbol{\alpha}_{//}$, parallel to $\mathbf{k}_l$, given by

$$\boldsymbol{\alpha}_{//} = (\omega/c)K \cos (\Psi_N - \Psi_K) (\sin \Psi_N, 0, \cos \Psi_N). \tag{3.3}$$

Referring to Fig. 1, we see that the following relation holds:

$$\boldsymbol{\alpha}_\perp + \boldsymbol{\alpha}_{//} = (\omega/c)K \cos (\Psi_N - \Psi_K) (-\cos \Psi_N + \sin \Psi_N, 0, \sin \Psi_N + \cos \Psi_N) = (\omega/c)K (\sin \Psi_K, 0, \cos \Psi_K). \tag{3.4}$$

Using the definitions of $N$, $K$, $\Psi_N$, and $\Psi_K$ expressed in terms of $n_l$ and $\kappa$ given in Sec. II, and noting that $\mathbf{k}_l \cdot \boldsymbol{\alpha}_\perp = 0$ and $\boldsymbol{\alpha}_\perp \cdot \boldsymbol{\alpha}_{//} = 0$, we have

$$[\mathbf{k}_l - j(\boldsymbol{\alpha}_\perp + \boldsymbol{\alpha}_{//})] \cdot [\mathbf{k}_l - j(\boldsymbol{\alpha}_\perp + \boldsymbol{\alpha}_{//})] = (\omega/c)^2[N^2 - K^2 - 2jNK \cos (\Psi_N - \Psi_K)]$$

$$= (\omega/c)^2(n_l^2 - \kappa^2 - 2jn_l\kappa). \tag{3.5}$$

Equating the real and imaginary parts, we get a set of equations due to Ketteler [22, 23]:

$$N^2 - K^2 = n_l^2 - \kappa^2 \tag{3.6}$$

and

$$NK \cos (\Psi_N - \Psi_K) = n_l\kappa. \tag{3.7}$$

Thus $N$ and $K$ are not independent but connected through (3.6) and (3.7).

Examining (3.7) above, we see there emerge two distinct cases of particular interest. First, inserting $\Psi_N = \pi/2$ in (3.7) reveals that, in the presence of gain or loss ($n_l \neq 0$ and $\kappa \neq 0$), the $x$-component of the amplitude vector cannot be zero: $NK \sin \Psi_K \neq 0$ or $\boldsymbol{\alpha}_\perp \neq \boldsymbol{0}$. This makes perfect physical sense in that, if there is an evanescent wave, it must get either amplified or attenuated as it propagates in the $x$-direction. On the other hand, if there is no amplitude change along the $x$-direction, that is, $\sin \Psi_K = 0$ or $\boldsymbol{\alpha}_{//} = \boldsymbol{0}$, the refraction angle $\Psi_N$ cannot reach $\pi/2$, hence no total reflection.

We must now discuss the derivation of the law of refraction since it will prove crucial for the treatment of total internal reflection from a complex-index medium. Reminding ourselves that the law of refraction is a manifestation of phase matching at the boundary, we require that the phase propagation vectors of the incident and refracted waves projected onto the boundary plane be equal to each other:

$$n_h \sin \theta = N \sin \Psi_N. \qquad (3.8)$$

This is the law of refraction when one of the two media in contact has a complex index of refraction. For a given set of $n_h$, $n_l$, and $\kappa$, we can determine the critical angle $\theta_c$ for total internal reflection by seeking the value of $\psi'$ which gives a minimum $\theta$ satisfying $n_h \sin \theta \geq N \sin (\pi/2) = (n_l^2 \cosh^2 \psi' + \kappa^2 \sinh^2 \psi')^{1/2}$.

A slight rearrangement of the right-hand side of the above

$$n_h \sin \theta \geq n_l \cosh \psi' \, [1+ (\kappa/n_l)^2 \tanh^2\psi']^{1/2}, \tag{3.9}$$

allows us to see that the effect of the index of extinction is merely to modify Snell's law for the transparent case $n_h \sin \theta \geq n_l \cosh \psi'$ by the factor $[1+ (\kappa/n_l)^2 \tanh^2\psi']^{1/2}$, the second term in the brackets being a correction arising from $\kappa$, which does not make itself felt in almost all the practical cases where $|\kappa|/n_l \ll 1$ holds [24].

In contrast to the derivation above, which is solely based on phase matching, the standard approach generally accepted in the literature, apparently, is to make use of the transparent-case expression $n_h \sin \theta = n_l \sin \psi$ with a formal substitution of the real quantities $n_l$ and $\psi$ by the complex ones $n_l - j\kappa$ and $\psi - j\psi'$, resulting in

$$n_h \sin \theta = (n_l - j\kappa) \sin (\psi - j\psi')$$

$$= N \sin \Psi_N - jK \sin \Psi_K. \tag{3.10}$$

The consequence of this simple-substitution approach immediately becomes apparent. With the left-hand side being purely real, one would end up with $\boldsymbol{\alpha}_{//} = \mathbf{0}$ or $K \sin \Psi_K = 0$. We have seen that in the preceding discussion that we cannot have $\Psi_N = \pi/2$ if the $x$-component of the amplitude vector vanishes: $K \sin \Psi_K = 0$. Consequently, taking the path of formal substitution amounts to imposing a physically unfounded constraint on $\Psi_N$ that forbids us to arrive at a solution corresponding to total reflection.

We finally proceed to discuss the directions of the transverse and longitudinal amplitude vectors $\boldsymbol{\alpha}_\perp$ and $\boldsymbol{\alpha}_{//}$. At the end of Sec. II, we were left with the following expressions:

$$[\boldsymbol{\alpha}_\perp + \boldsymbol{\alpha}_{//}]_x = K \sin \Psi_K = \kappa \sin \psi \cosh \psi' + n_l \cos \psi \sinh \psi' \tag{3.11}$$

and

$$[\boldsymbol{\alpha}_\perp + \boldsymbol{\alpha}_{//}]_z = K \cos \Psi_K = (+/-)(\kappa \cos \psi \cosh \psi' - n_l \sin \psi \sinh \psi'), \qquad (3.12)$$

where $K = (\kappa^2 \cosh^2\psi' + n_l^2 \sinh^2\psi')^{1/2}$. The sign on the right-hand side of (3.12) must be chosen in such a way that the following two requirements are met:

(i) For incidence below the critical angle $\theta < \theta_c$, the refracted field gets attenuated for $\kappa > 0$ and amplified for $\kappa < 0$ as it is going away from the boundary plane. In such cases, the planes of equal phase in the low-index medium are perpendicular to the direction of propagation whereas, since the amplitude change depends directly on the distance travelled in the medium, the loci of points of equal amplitude will be planes parallel to the surface of separation. Thus we have $[\boldsymbol{\alpha}_\perp + \boldsymbol{\alpha}_{//}]_x = (\omega/c)K \sin \Psi_K = 0$ for both $\kappa > 0$ and $\kappa < 0$. As for the z-direction, we must have $[\boldsymbol{\alpha}_\perp + \boldsymbol{\alpha}_{//}]_z = (\omega/c)K \cos \Psi_K < 0$ for $\kappa > 0$ and $[\boldsymbol{\alpha}_\perp + \boldsymbol{\alpha}_{//}]_z = (\omega/c)K \cos \Psi_K > 0$ for $\kappa < 0$.

(ii) Under total reflection ($\theta > \theta_c$), the fields are evanescent in the z-direction $[\boldsymbol{\alpha}_\perp + \boldsymbol{\alpha}_{//}]_z > 0$ for both $\kappa > 0$ and $\kappa < 0$.

In the following, we discuss four cases, where the incidence is in either the below-the-critical-angle ($\theta < \theta_c$) or the total-reflection ($\theta \geq \theta_c$) region, for each of which there are two cases: $\kappa > 0$ for loss and $\kappa < 0$ for gain.

**(a) $\theta < \theta_c$ and $\kappa > 0$**

In the region $\theta < \theta_c$, the amplitude vector is normal to the boundary plane [Fig. 2(a)]. Thus we have $[\boldsymbol{\alpha}_\perp + \boldsymbol{\alpha}_{//}]_x = (\omega/c)K \sin \Psi_K = (\omega/c)(\kappa \sin \psi \cosh \psi' + n_l \cos \psi \sinh \psi') = 0$, from which

we have

$$\tan\psi/\tanh\psi' = -n_l/\kappa.$$

With both $n_l$ and $\kappa$ positive and $0 \leq \psi < \pi/2$ in the above, we see that $\psi' < 0$ [Fig. 2(a)]. For the refracted field to attenuate in the positive z-direction, we must have $[\alpha_\perp + \alpha_{//}]_z = (\omega/c)K \cos\Psi_K > 0$, which leads us to take the plus sign in (3.12):

$$K\cos\Psi_K = \kappa \cos\psi \cosh\psi' - n_l \sin\psi \sinh\psi'.$$

**(b) $\theta < \theta_c$ and $\kappa < 0$**

Just as we have done with the attenuating case above, we start with

$$\tan\psi/\tanh\psi' = -n_l/\kappa.$$

With both $n_l > 0$ and $\kappa < 0$ and $0 \leq \psi < \pi/2$, we have $\psi' > 0$ in the above. The negative $\kappa$ means an exponential growth in the z-direction [Fig. 2(b)], hence $[\alpha_\perp + \alpha_{//}]_z = (\omega/c)K \cos\Psi_K < 0$, which leads us to take the plus sign in (3.12):

$$K\cos\Psi_K = \kappa \cos\psi \cosh\psi' - n_l \sin\psi \sinh\psi'.$$

**(c) $\theta \geq \theta_c$ and $\kappa > 0$**

In the total-reflection region where $\Psi_N = \pi/2$, the phase propagation vector is parallel to the x-axis, $[k_l]_z = (\omega/c)N \cos\Psi_N = (\omega/c)(n_l \cos\psi \cosh\psi' + \kappa \sin\psi \sinh\psi') = 0$ [Fig. 2(c)], from which we get

$$\tan\psi \tanh\psi' = -n_l/\kappa,$$

where $n_l > 0$ and $\kappa > 0$. Recognising that $\Psi_N \sim \psi$ when $n_l \gg \kappa$, as is most often with the practical cases, we can assume that $\psi$ lies somewhere between 0 and $\pi/2$ just as $\Psi_N$ does.

From tan $\psi$ tanh $\psi'$ = $-n_l/\kappa$ < 0 with 0 ≤ $\psi$ < $\pi/2$, we get $\psi'$ < 0. By choosing the plus sign in (3.12):

$$K \cos \Psi_K = \kappa \cos \psi \cosh \psi' - n_l \sin \psi \sinh \psi',$$

we have at all angles of incidence above $\theta_c$, $[\boldsymbol{\alpha}_\perp + \boldsymbol{\alpha}_{//}]_z = K \cos \Psi_K > 0$, which ensures the refracted field to be evanescent.

**(d) $\theta \geq \theta_c$ and $\kappa < 0$**

The same argument applies as the case (c) above with a difference being $\kappa < 0$ giving $\psi' > 0$. For the refracted fields to be evanescent, $[\boldsymbol{\alpha}_\perp + \boldsymbol{\alpha}_{//}]_z = K \cos \Psi_K > 0$ [Fig. 2(d)], we choose the minus sign in (3.12):

$$K \cos \Psi_K = - \kappa \cos \psi \cosh \psi' + n_l \sin \psi \sinh \psi'.$$

## IV. Phase reflection coefficient and the Goos-Hänchen shift

Having transformed the physical parameters such as $n_l$ and $\kappa$ into the effective indices of refraction and extinction, $N$ and $K$, associated with the angles, $\Psi_N$ and $\Psi_K$, which allow us to define the complex propagation vectors, we are ready to calculate the phase and amplitude reflection coefficients.

We assume that the refracted electric field takes the form:

$$E_y'' (x, z, t) = E_s'' \exp[-j(\boldsymbol{k}_l - j\boldsymbol{\alpha}_\perp)\cdot\boldsymbol{r}] \exp(j\omega t), \qquad (4.1)$$

where $\boldsymbol{k}_l = ((\omega/c)N \sin \Psi_N, 0, 0)$, $\boldsymbol{\alpha}_\perp = (0, 0, (\omega/c)K \cos \Psi_K)$, and $\boldsymbol{r} = (x, y, z)$. A caution is

appropriate here: if a factor such as $\exp(-\boldsymbol{\alpha}_{//}\cdot\boldsymbol{r})$ [$= \exp(-(\omega/c)N \sin \Psi_N\, x)$] appeared on the right-hand side, we would end up with an unphysical solution, for which the amplitude of the evanescent field keeps either decreasing ($\kappa > 0$) or increasing ($\kappa < 0$) along the $x$ axis. In order to accommodate such a solution that remain constant with the coordinate $x$ even with gain or loss, we shall let the $\boldsymbol{\alpha}_{//}$-dependence be absorbed into the constant amplitude $E_s''$ so that $E_s''$ not be an explicit function of the coordinate $x$.

By use of the Maxwell equations satisfied by the $s$-polarised waves, we find that the corresponding magnetic field has components of the form:

$$H_x''(x, z, t) = (jk_0 K/\mu_0\omega) \cos \Psi_K\, E_s''\, \exp[-j(\boldsymbol{k}_l - j\boldsymbol{\alpha}_\perp)\cdot\boldsymbol{r}] \exp(j\omega t) \qquad (4.2)$$

and

$$H_z''(x, z, t) = (-k_0 N/\mu_0\omega) \sin \Psi_N\, E_s''\, \exp[-j(\boldsymbol{k}_l - j\boldsymbol{\alpha}_\perp)\cdot\boldsymbol{r}] \exp(j\omega t). \qquad (4.3)$$

Our task now is to determine the refracted fields $E_s''$ and $H_s''$ along with the reflection coefficient so that we could express the reflected field $E_s'$ as a function of the angle of incidence. The reflection coefficient is given by a product of coefficients accounting for amplitude and phase changes:

$$r_s = \rho_s \exp(j\Phi_s). \qquad (4.4)$$

Here $\rho_s$ denotes the amplitude reflection coefficient, which is defined as $\rho_s = |E_s'|/|E_s|$, and $\Phi_s$ is the phase change that occurs upon total reflection. The transparent-case reflection coefficient $r_{s0}$, the limiting case of (4.4) for which the amplitude reflection coefficient is unity, is given by

$$r_{s0} = \exp(j\Phi_{s0}) \tag{4.5}$$

with $\Phi_{s0} = 2\arctan[n_l \sinh \psi'/n_h \cos\theta] = 2\arctan[(n_h^2 \sin^2\theta - n_l^2)^{1/2}/n_h \cos\theta]$.

The phase shift undergoes a slight modification due to the index of extinction. By inspection we recognise that the complex-index case analogue is obtained if we replace the lossless-case expression $[\boldsymbol{\alpha}_{\perp 0}]_z = (\omega/c)n_l \sinh\psi' = (\omega/c)(n_h^2\sin^2\theta - n_l^2)^{1/2}$ by $[\boldsymbol{\alpha}_\perp]_z = (\omega/c)K\cos\Psi_K$. The phase change for $s$-polarised waves becomes

$$\Phi_s = 2\arctan(K\cos\Psi_K/n_h \cos\theta). \tag{4.6}$$

As we have seen in the discussion at the end of Sec. III, $K\cos\Psi_K > 0$ is satisfied in either case of $\kappa > 0$ or $\kappa < 0$. Note that the transparent case (4.5) is recovered in the limit of $\kappa$ approaching zero.

Having incorporated to the phase change $\Phi_s$ the correction due to $\kappa$, we now address ourselves to finding an expression for the Goos-Hänchen shift. The transparent-case shift for $s$-polarised waves is given by [25]

$$\delta_{s0} = 2\cos\theta \sin\theta \, [1 - n_l^2/n_h^2]^{-1} d_{s0}, \tag{4.7}$$

where $d_{s0} = (c/\omega)(n_h^2\sin^2\theta - n_l^2)^{-1/2} = [(\omega/c) n_l \sinh\psi']^{-1}$. In the preceding paragraph, we have observed that the complex-index analogues can be obtained by replacing $n_l \sinh\psi'$ by $K\cos\Psi_K$. Thus the penetration depth becomes $d_s = [\boldsymbol{\alpha}_\perp]_z^{-1} = [(\omega/c) K\cos\Psi_K]^{-1}$.

Now $n_l^2$ in the denominator of (4.6) is to be modified accordingly. Allowing for a development parallel to the transparent case $(\boldsymbol{k}_{l0} - j\boldsymbol{\alpha}_{\perp 0})\cdot(\boldsymbol{k}_{l0} - j\boldsymbol{\alpha}_{\perp 0}) = (\omega/c)^2 n_l^2$ we found in Sec. II, we are led to $(\mathbf{k}_l - j\boldsymbol{\alpha}_\perp)\cdot(\mathbf{k}_l - j\boldsymbol{\alpha}_\perp) = (\omega/c)^2[N^2 \sin^2(\pi/2) - K^2 \cos^2 \Psi_K]$ for $\kappa \neq 0$. We observe that a formal replacement of $n_l^2$ by $N^2 - K^2 \cos^2 \Psi_K$ brings our complex-index case into correspondence with the transparent case. Thus we find for the Goos-Hänchen shift for the complex-index case

$$\delta_s = 2\cos\theta \sin\theta \, [1 - (N^2 - K^2 \cos^2 \Psi_K)/n_h^2]^{-1} d_s, \qquad (4.8)$$

where $d_s = [(\omega/c)K \cos \Psi_K]^{-1}$. Naturally, the transparent case (4.6) is recovered for $\kappa = 0$.

## V. Amplitude reflection coefficient

The present section deals with the problem of finding the amplitude reflection coefficient. The discussion on the derivation of Snell's law in Sec. III suggests that, to obtain correct expressions for total reflection from an amplifying or attenuating medium, we cannot naively rely on replacing in a purely formal way the real quantities such as $n_l$ and $\psi$ in the Fresnel formulas by complex ones $n_l - j\kappa$ and $\psi - j\psi'$. Instead we shall resort to formulating the Helmholtz equation satisfied by the amplitude reflection coefficient $\rho_s$. By solving it, explicit expressions are obtained for the electric and magnetic fields of the refracted wave.

To find the differential equation satisfied by $\rho_s$, we first express the refracted field $E_y''$ in terms of $\rho_s$. The evanescent wave does not exist on its own; it always is associated with a homogeneous wave in the higher-index medium. Thus the refracted field is given by the superposition of two evanescent fields, one which is associated with the incident field and the other with the reflected field:

$$E_y'' = E_y'' \text{ (incident)} + E_y'' \text{ (reflected)}, \tag{5.1}$$

where

$$E_y'' \text{ (incident)} = E_s \exp[-j(\mathbf{k}_l - j\boldsymbol{\alpha}_\perp)\cdot\mathbf{r}] \tag{5.2}$$

and

$$E_y'' \text{ (reflected)} = \rho_s \exp(j\Phi_s) E_s \exp[-j(\mathbf{k}_l - j\boldsymbol{\alpha}_\perp)\cdot\mathbf{r}]. \tag{5.3}$$

Note that, as discussed in the beginning of Sec. IV, we lump into $\rho_s$ the $\boldsymbol{\alpha}_{//}$-dependence, which accounts for amplitude change in the $\mathbf{k}_l$-direction. We can readily see that (5.2) is a solution to

$$\nabla^2 E_y'' \text{ (incident)} = -(\mathbf{k}_l - j\boldsymbol{\alpha}_\perp)^2 E_y'' \text{ (incident)}. \tag{5.4}$$

To determine the amplitude reflection coefficient, we need to solve the Helmholtz equation satisfied by the evanescent field associated with the reflected field (5.3):

$$\nabla^2 E_y'' \text{ (reflected)} = -[\mathbf{k}_l - j(\boldsymbol{\alpha}_\perp + \boldsymbol{\alpha}_{//})]^2 E_y'' \text{ (reflected)}, \tag{5.5}$$

from which, after cancellations and rearrangements with the help of (5.2) and (5.4), we arrive at a differential equation for $\rho_s$:

$$d^2\rho_s/dx^2 - 2j(\omega/c)N \sin \Psi_N\, d\rho_s/dx - (\omega/c)^2 N^2 \sin^2 \Psi_N\, \rho_s + (\omega/c)^2 K^2 \cos^2 \Psi_K\, \rho_s = -(\omega/c)^2 [N^2 - K^2 -$$
$$2jNK \cos(\Psi_N - \Psi_K)]\, \rho_s. \tag{5.6}$$

Equating the real terms, we obtain

$$d^2\rho_s/dx^2 - (\omega/c)^2 K^2 \sin^2 \Psi_K\, \rho_s = 0,$$

or

$$[d/dx + (\omega/c)K \sin \Psi_K]\, \rho_s = 0 \text{ or } [d/dx - (\omega/c)K \sin \Psi_K]\, \rho_s = 0. \tag{5.7}$$

Here we employ a model in which a given point on the wavefront goes into the low-index medium at $x = 0$, traverses a distance equal to the Goos-Hänchen shift $\delta_s$, experiencing gain or loss, and then emerges into the high-index medium [20, 25]. Thus, noting that $\rho_s = 1$ at $x = 0$, we may integrate the first one of (5.7) above over the propagation distance $\delta_s$ to find

$$\rho_s = \exp[-(\omega/c)K\sin\Psi_k\delta_s]. \tag{5.8}$$

The second equation of (5.7) gives an unphysical solution and is discarded here. Also, equating the imaginary terms in (5.6) may as well yield (5.8).

As we have seen at the end of Sec. III, the longitudinal amplitude vector $\boldsymbol{\alpha}_{/\!/}$ points in the positive $x$-direction for $\kappa > 0$ and the negative $x$-direction for $\kappa < 0$, leading to $\rho_s < 1$ for $\kappa > 0$ and $\rho_s > 1$ for $\kappa < 0$. Thus we have established that Maxwell's theory allows for total reflection with a greater-than-unity amplitude reflection from a gain medium. When $n_l \gg |\kappa|$, implying $\psi = \pi/2$ and $\psi' \sim 0$, (5.8) reduces to $\rho_s = \exp[-(\omega/c)\kappa\delta_{s0}]$ with $\delta_{s0}$ given in (4.7). This is the expression obtained in [20] through a different route.

Substituting (4.8) and (5.8) into (5.2) and (5.3), and noting that $K\cos\Psi_K/n_h\cos\theta = \tan(\Phi_s/2)$, we have for the electric field written out in full

$$E_y'' = E_s'' \exp(-j\boldsymbol{k}_l\cdot\boldsymbol{r})\exp(-\boldsymbol{\alpha}_\perp\cdot\boldsymbol{r}). \tag{5.9}$$

where

$$E_s'' = t_s E_s \{[\exp(-(\omega/c)K\sin\Psi_K\delta_s) + 1] + j\tan(\Phi_s/2)[\exp(-(\omega/c)K\sin\Psi_K\delta_s) - 1]\}.$$

with $t_s = \cos(\Phi_s/2)\exp(j\Phi_s/2)$. By use of the Maxwell equations, the corresponding $x$- and $z$-components of the magnetic field immediately follow:

$$H_x'' = -jK \cos \Psi_K (\varepsilon_0/\mu_0)^{1/2} E_s'' \exp(-j\mathbf{k}_l\cdot\mathbf{r}) \exp(-\boldsymbol{\alpha}_\perp\cdot\mathbf{r}) \qquad (5.10)$$

and

$$H_z'' = N \sin \Psi_N (\varepsilon_0/\mu_0)^{1/2} E_s'' \exp(-j\mathbf{k}_l\cdot\mathbf{r}) \exp(-\boldsymbol{\alpha}_\perp\cdot\mathbf{r}), \qquad (5.11)$$

where $E_s''$ is given in (5.9).

Notice that the planes of constant amplitude, being perpendicular to $\boldsymbol{\alpha}_\perp = (0, 0, (\omega/c) \sin \Psi_K)$, are parallel to the boundary plane. The refracted fields consist of a part that is in-phase and the other that is in-quadrature with the incident field. In the transparent case ($\kappa = 0$), where $\boldsymbol{\alpha}_{//}$ vanishes, and $\rho_s$ becomes unity, the in-phase term reduces to the known transparent case expression while the in-quadrature term vanishes.

## VI. Conclusions

We have considered the total reflection of a plane electromagnetic wave at a boundary separating two media, one of which is transparent and the other, lower-index one is either amplifying or attenuating. The amplification and attenuation of evanescent fields have received separate treatments historically, as we have seen in the introduction. Those problems have been merged into one and solved in a unified manner from the standpoint of classical electrodynamics. This does make physical sense, for the problem of amplification or attenuation should be simply a matter of reversing the sign of the index of extinction $\kappa$ in the domain of classical theory.

The essence of the treatment lies in judiciously formulating the law of refraction, leading us to plane-wave solutions to Maxwell's equations, which describe the amplification or attenuation of the evanescent fields through their interaction with the totally reflecting medium. The analytical expressions have been obtained for the phase and amplitude reflection coefficients as well as the electric and magnetic components of the evanescent fields. Those are given in terms of the effective indices of refraction and extinction associated with the real angles of the phase and amplitude change. For the sufficiently small ratio $\kappa/n_l$, the expressions obtained here reduce to those deduced by means of a different method.

## Appendix: *p*-polarisation

Here we shall derive expressions for the *p*-polarisation case, tracing the same line of discussion as the *s*-polarisation case with the only difference being somewhat more involved in places. For *p*-polarised waves, we deal with $E_x''$, $E_z''$, and $H_y''$, instead of $E_y''$, $H_x''$, and $H_z''$ (Fig. 1). Starting with the phase reflection coefficient for transparent case

$$r_{p0} = \exp(j\Phi_{p0}),$$

where $\Phi_{p0} = 2\arctan\{(n_l/n_h)^2[(n_h^2\sin^2\theta - n_l^2)^{1/2}/n_h\cos\theta]\}$. As was done with the *s*-polarisation case in Sec. IV, its complex-index analogues is found by replacing $n_l^2$ by $N^2 - K^2\cos^2\Psi_K$. Thus we find for *p*-polarised waves

$$\Phi_p = 2\arctan\{[(N^2 - K^2\cos^2\Psi_K)/n_h^2](K\cos\Psi_K/n_h\cos\theta)\}. \tag{A.1}$$

As was done for s-polarised waves in Sec V, we treat the evanescent magnetic fields which are excited by the incident and reflected fields, $H_y''$ (incident) and $H_y''$ (reflected), separately. The Helmholtz equation for $H_y''$ (incident) is

$$\nabla^2 H_y'' \text{ (incident)} = -(k_l - j\boldsymbol{\alpha}_\perp)^2 H_y'' \text{ (incident)}, \quad (A.2)$$

from which we obtain

$$H_y'' \text{ (incident)} = (\varepsilon_0/\mu_0)^{1/2} (N^2 - K^2 \cos^2 \Psi_K)^{1/2} t_p E_p \exp(-j\boldsymbol{k}_l \cdot \boldsymbol{r}) \exp(-\boldsymbol{\alpha}_\perp \cdot \boldsymbol{r}), \quad (A.3)$$

where $t_p = \cos(\Phi_p/2) \exp(j\Phi_p/2)$. The evanescent field associated with the reflected field

$$\rho_p \exp(j\Phi_p) (\varepsilon_0/\mu_0)^{1/2} (N^2 - K^2 \cos^2 \Psi_K)^{1/2} t_p E_p \exp(-j\boldsymbol{k}_l \cdot \boldsymbol{r}) \exp(-\boldsymbol{\alpha}_\perp \cdot \boldsymbol{r}), \quad (A.4)$$

satisfies the Helmholtz equation

$$\nabla^2 H_y'' \text{ (reflected)} = -[k_l - j(\boldsymbol{\alpha}_\perp + \boldsymbol{\alpha}_{//})]^2 H_y'' \text{ (reflected)}, \quad (A.5)$$

from which we obtain the amplitude reflection coefficient

$$\rho_p = \exp[-(\omega/c) K \sin \Psi_k \delta_p], \quad (A.6)$$

with the Goos-Hänchen shift for $p$-polarised waves $\delta_p$. The conversion of expressions between $s$- and $p$-polarisation can be done by use of what is termed the reduction factor $q_0$ given by [26]

$$q_0 = (n_h \sin \theta/n_l)^2 + (n_h \sin \theta/n_h)^2 - 1.$$

Just as we did with the phase change above, we can obtain its complex-index analogue through the substitution of $n_l^2$ by $N^2 - K^2 \cos^2\Psi_K$, that is

$$q = (N^2 - K^2 \cos^2\Psi_K)^{-1} (N^2 \sin^2\theta + K^2 \cos^2\Psi_K \cos^2\theta). \quad (A.7)$$

By use of the reduction factor $q$, the penetration depth for $p$-polarised waves is related with that for $s$-polarised waves $d_s$ by

$$d_p = q d_s, \quad (A.8)$$

where $d_s = (k_0 K \cos \Psi_K)^{-1}$. Finding that $\delta_p$ is written as

$$\delta_p = 2\cos\theta \sin\theta \, [1 - (N^2 \sin^2 \Psi_N - K^2 \cos^2 \Psi_K)/n_h^2]^{-1} d_p, \qquad (A.9)$$

we obtain

$$H_y'' = (\varepsilon_0/\mu_0)^{1/2} (N^2 - K^2 \cos^2 \Psi_K)^{1/2} E_p'' \exp(-j\mathbf{k}_l\cdot\mathbf{r}) \exp(-\boldsymbol{\alpha}_\perp\cdot\mathbf{r}), \qquad (A.10)$$

where

$$E_p'' = t_p \, E_p \, \{[\exp(-(\omega/c)K \sin \Psi_K \, \delta_p) + 1] + j\tan(\Phi_p/2)\,[\exp(-(\omega/c)K \sin \Psi_K \, \delta_p) - 1]\}.$$

with $t_p = \cos(\Phi_p/2) \exp(j\Phi_p/2)$.

By use of the Maxwell equations satisfied by the $p$-polarised components, we find the electric components of the fields:

$$E_x'' = -jK \cos \Psi_K (N^2 - K^2 \cos^2 \Psi_K)^{-1/2} E_p'' \exp(-j\mathbf{k}_l\cdot\mathbf{r}) \exp(-\boldsymbol{\alpha}_\perp\cdot\mathbf{r}), \qquad (A.11)$$

and

$$E_z'' = (N \sin \Psi_N)(N^2 - K^2 \cos^2 \Psi_K)^{-1/2} E_p'' \exp(-j\mathbf{k}_l\cdot\mathbf{r}) \exp(-\boldsymbol{\alpha}_\perp\cdot\mathbf{r}), \qquad (A.12)$$

respectively.

**Figure captions**

Fig. 1. Electric and magnetic fields of the incident, reflected, and refracted waves at the boundary between two media. The electric fields point out of the page in the *s*-polarisation case whereas the electric fields are in the polarisation plane in the *p*-polarisation case. In the transparent case ($\kappa = 0$), $\Psi_N$ reduces to $\psi$. Note that the drawing is just for the sake of illustrating the notation and that the situation depicted here is physically unrealisable, as one can see from the discussions in Sec. III.

Fig. 2. Schematic illustration of the phase propagation and amplitude vectors for below- [(a) and (b)] and above-the-critical-angle [(c) and (d)] incidence with the reflecting medium having gain [(b) and (d)] or loss [(a) and (c)]. Note that $\boldsymbol{\alpha}_{//} = \mathbf{0}$ for $\theta < \theta_c$ [(a) and (b)] and $\boldsymbol{\alpha}_{//} \neq \mathbf{0}$ for $\theta \geq \theta_c$ [(c) and (d)].

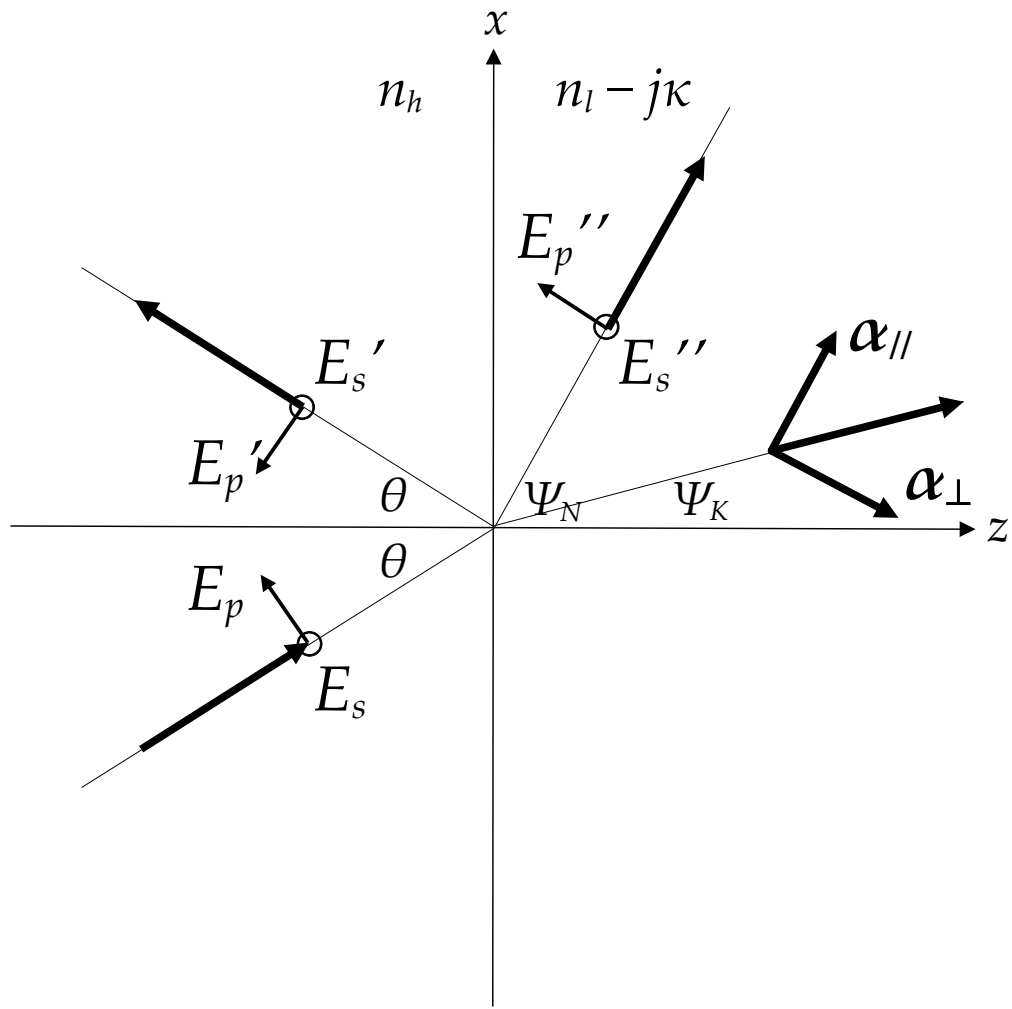

**Fig. 1** T. Kobayashi

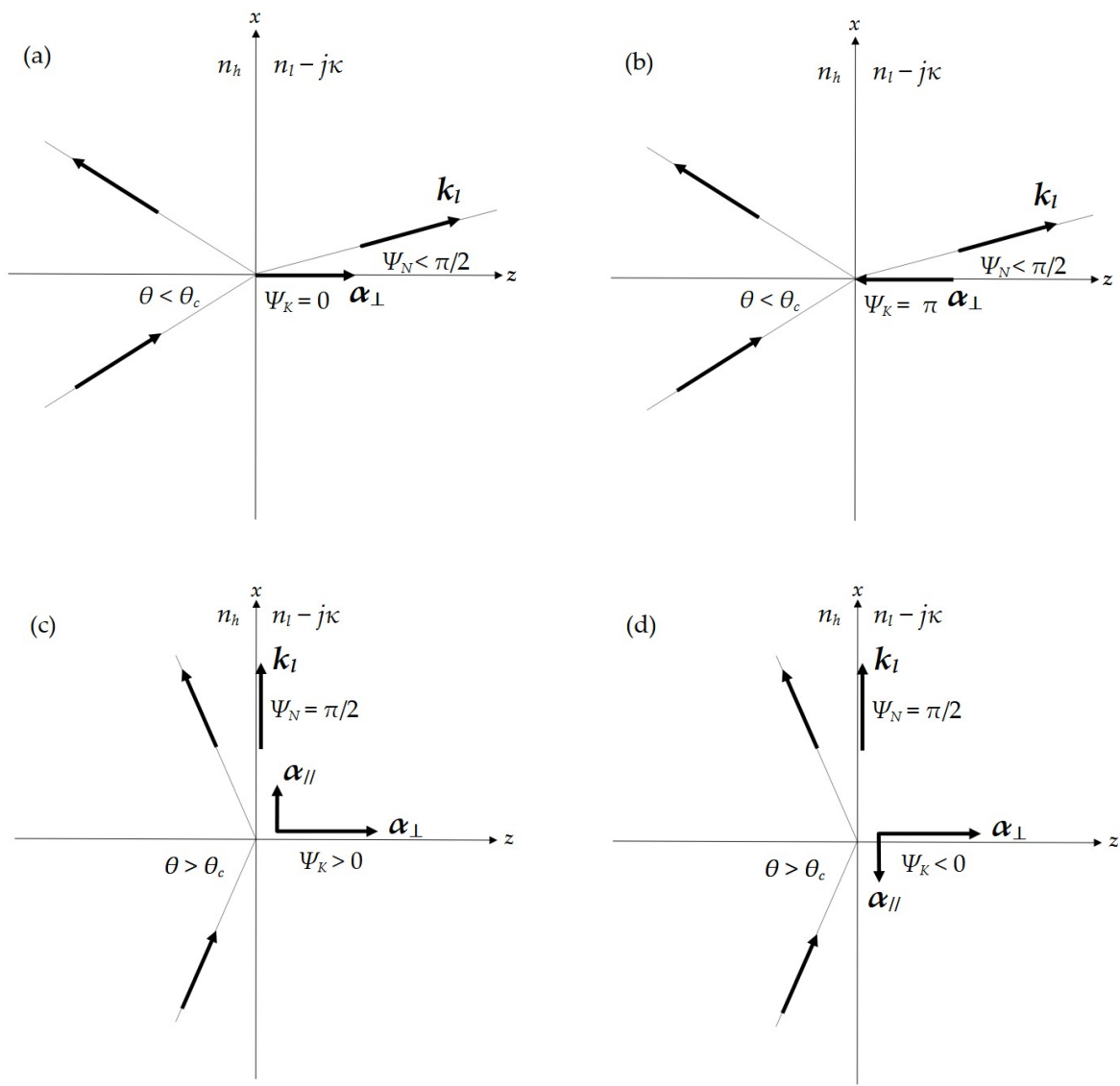

**Fig. 2 T. Kobayashi**